\documentstyle[11pt]{article}
\begin{document}
\oddsidemargin=-1cm
\topmargin=-1.5cm
\begin{center} {\Large {\bf Nonequilibrium Phase Transitions in 
Model Ferromagnets: A Review}}\end{center}

\begin{center} {\it Muktish Acharyya} \end{center}

\begin{center} {Department of Physics, Krishnanagar Government College,\\
 PO-Krishnanagar, Dist-Nadia, PIN-741101, West-Bengal, India.\\
 {\it E-mail:muktish@vsnl.net}}\end{center}

The thermodynamical behaviours of ferromagnetic systems in equilibrium 
are well studied. However, the ferromagnetic system, far from equilibrium, became an
interesting field of research in last few decades.
The ferromagnetic systems in the presence of a steady magnetic field are also studied
by using standard tools of equilibrium statistical physics. The ferromagnet in the 
presence of time-dependent magnetic field, shows various interesting phenomena, 
explored recently. An usual response of a ferromagnet in the presence of a
sinusoidally oscillating magnetic field is the hysteresis. Apart from this hysteretic
response, the nonequilibrium dynamic phase transition is a very interesting phenomenon.
In this chapter, the nonequilibrium dynamic phase transitions, in model ferromagnetic 
systems in presence of time-dependent magnetic field, are discussed. 
For this kind of nonequilibrium phase transition one cannot 
employ the standard techniques of equilibrium statistical mechanics.  The recent developments 
in this direction are mainly based on numerical simulation (Monte Carlo). The Monte Carlo 
simulation, of kinetic Ising model in presence of sinusoidally oscillating 
(in time but uniform over space) magnetic field, 
is extensively performed to study the nonequilibrium dynamic phase transition.  The temperature 
variations of dynamic order parameter, dynamic specific heat, dynamic relaxation time etc. 
near the transition point are discussed. The appearance of a dynamic length scale  
and a dynamic time scale and their 
behaviours near the transition point are also discussed. All these studies indicate that this 
proposed dynamic transition is a nonequilibrium thermodynamic phase transition. 
The disorder (quenched) induced zero temperature (athermal) dynamic transition is studied in
random field Ising ferromagnet.
The dynamic transition in the 
Heisenberg ferromagnet is also studied. The nature of this transition in the Heisenberg ferromagnet
depends on the anisotropy and the polarisation of the applied time varying magnetic field. 
The anisotropic Heisenberg ferromagnet in the presence of elliptically polarised magnetic field 
shows multiple dynamic transitions. This multiple dynamic transitions in anisotropic Heisenberg
ferromagnet are discussed here. 
Recent experimental evidences of dynamic transitions are also discussed very briefly.

\vskip 1 cm


\vskip 1 cm

\noindent {\large {\bf 1. Introduction:}}

The ferromagnetic system in statistical equilibrium gives rise to well-known ferro-para phase transition. However,
the ferromagnetic system, in the presence of a time varying external magnetic field, remains
far from thermodynamical/statistical equilibrium. This type of system became an
interesting object of research over the last two decades. Particularly, if the time varying externally
applied magnetic field is sinusoidally oscillating (in time but uniform over the space), it yields
two major responses of the ferromagnetic systems, i.e. (i) dynamic hysteresis (ii) dynamic phase
transition. The dynamic hysteretic responses of Ising-like systems are already discussed in a recent
review article \cite{rev}. But recently, the another response, i.e., the dynamic transition
has drawn much attention of researchers. It has several interesting and important (in the context of
nonequilibrium phase transition) aspects and can provide a simple example of nonequilibrium transition.
In the earlier review \cite{rev}, the dynamic transition was also discussed in brief. However, in last
five years a considerable amount of research was done on it (in the Ising model) which was not reviewed yet. 
Very recently, the study of the dynamic transitions in the Heisenberg model gives several interesting
features which are not possible to observe in the Ising models.  
Keeping this in mind, I have tried to review the recent developements in this field. 

This review is organised as follows: in the next section a short introduction is written about what a dynamic
transition is and when it should be called true dynamic transition. In third section, the dynamic transition
in the Ising ferromagnet is discussed. 
Different aspects of 
dynamic transitions (mainly its thermodynamic natures) and its relationship with the stochastic resonance
are reviewed. 
In this review the analytic formulation of dynamic transition and the {\it athermal}
dynamic transition. 
In the fourth section, the dynamic transition in the Heisenberg ferromagnet is discussed. This section gives
a review of very recent (last 1-2 years) works on classical vector spin models. The results
available so far, based on the experimental observations reviewed in the fifth section. The chapter ends with
a summary given in the sixth section. 

\vskip 1 cm

\noindent {\large {\bf 2. What is dynamic transition ?}

If the temperature of a ferromagnetic sample increases (in absence of external magnetic field), 
the sponteneous magnetisation vanishes
at a particular temperature (Curie temperature) and the transition occurs from an asymmetric
(ferromagnetic) phase to a symmetric (paramagnetic) phase. This is a very well known phenomenon
and the equilibrium symmetry breaking ferro-para transition is also well studied. But any nonequilibrium 
transition of any ferromagnetic sample in the presence of time-dependent 
magnetic field was unexplored even before 1990!! Tome
and Oliviera \cite{tom} first observed a prototype of nonequilibrium dynamic transition in the
numerical solution (by fourth order Runge-Kutta method) of mean field equation of motion 
\begin{equation}
{dm \over dt} = -m + {\rm tanh}{\left[{{m(t)+h(t)} \over T}\right]}
\label{mf}
\end{equation}
of the classical one component ($m$) ferromagnetic  
model in presence of a magnetic field varying sinusoidally ($h(t) = h_0 {\rm cos}(\omega t)$) in time.  
Here, the time-averaged magnetisation $Q =  \oint m(t) dt/\tau$ ($\tau = 2\pi/\omega$ is the time period
of the oscillating field) over a full cycle of the oscillating field plays the role of order parameter
for this type of proposed nonequilibrium dynamic transition. Tome
and Oliveira \cite{tom} found that $Q$ becomes zero (dynamically disordered phase) from a nonzero 
(dynamically ordered phase) value at a finite temperature $T$ (which also depends
on the value of amplitude $h_0$ of oscillating field). They sketched a dynamic phase boundary in $h_0 - T$
plane. For the higher values of $h_0$ the transition was found to be of discontinuous type and that becomes continuous
for lower values of $h_0$. They \cite{tom} also located a tricitical point (TCP) on the phase boundary which separates the nature
(discontinuous/continuous) of these transitions.

This phenomenon can be explained by considering the system initially kept in one well of a Landau type 
double well potential. Depending on
the temperature, a definite amount of magnetic field is necessary to bring the system from one well to another.
If the amplitude of the applied oscillating magnetic field is less than the required amount, the system oscillates in one well (where
it was initially). In this situation, the magnetisation does not change its sign. As a result, the magnitude of
the time-averaged magnetisation is nonzero ($Q \ne 0$). If one sees the plot of $m(t)-h(t)$, it is asymmetric
about the $m(t) = 0$ line. This gives rise to a dynamically ordered and asymmetric phase. As the temperature increases,
the height of the barrier between two wells decreases and to push the system from one well to another, less amount
of magnetic field is necessary. As a result, the magnetisation can change its sign for this amount of field. 
Consequently, the time-averaged
magnetisation over a full cycle becomes zero ($Q = 0$) when $m(t)-h(t)$ plot is symmetric. So, we get the symmetry
broken dynamic phase transition, where the transition temperature decreases as the value of magnetic field
increases. This can qualitatively explain the phase boundary of dynamic phase transition.
Fig. \ref{dpt} illustrates this symmetry breaking associated with the dynamic transition.

But, can we really call this transition a true dynamic transition ? A true dynamic transition will be such that this transition
should disappear in the static limit !! Let us examine logically, what happens in the static (infinitesimally small frequency)
limit. One can simulate this situation by varying the magnetic field with infinitesimally small frequency. 
It should be noted here (in eqn.\ref{mf}) that the
dynamics of the one component magnetisation ($m$) is purely deterministic. So, at a particular temperature,
to bring the system from one well to another, a definite amount of field should be applied,
irrespective of the rate of achieving the required amount. The system will wait
untill it gets the required value of the field. If the frequency is very small, it will wait for very long time.
But unless it gets the required value of field it will not go to another well. There is no stochasticity involved
or any noise is present in the equation of motion which can push the system towards another well, irrespective of value
of the applied field. So, obviously, in the zero
frequency limit a so called dynamic transition will be observed in the case of meanfield study. Just by this argument 
one can immediately conclude that the kind of transition described above cannot be truely dynamic in nature!!

After realising this, the researchers, interested in this field,
 tried to study the true dynamic transition in ferromagnetic model systems
incorporating the thermal fluctuations as the source of noise or stochasticity, 
keeping in mind that this stochasticity will help to push the
system into another well. Rao, Krishnamurthy and Pandit \cite{rao} and Dhar and Thomas \cite{dd} 
tried to observe this perfectly dynamic 
transition in $N$-vector model in the $N \to \infty$ limit. They \cite{rao,dd} also tried to observe this in the
kinetic Ising model by Monte Carlo (MC) simulation. At the same time, Lo and Pelcovits \cite{lo} studied the kinetic
Ising model in the presence of sinusoidally oscillating magnetic field by Monte Carlo simulation. But, unfortunately
both  failed to observe the dynamic transition and to draw the phase boundary. Being motivated from these studies;
extensive MC studies \cite{dd,ac,ma1,ma2,ma3,ma4,ma5,ma6,rik1,rik2,rik3} were performed in the 
kinetic Ising model in oscillating magnetic field 
and the recent developments in this direction are discussed in details in the next section. It should be mentioned
here that the dynamic response at very low frequency, essentially originates from traditional nucleation problem \cite{stauffer}.
However, to see this effect one should use a frequency which is much smaller than those usually employed in this
field of research \cite{rev}.

Although the numerical solution of mean-field equation (Eqn. \ref{mf}) cannot provide for a 
true dynamic transition, it has one importance. The 
linearised equation is exactly solvable and can be used to get some qualitative features of dynamic transition analytically. 

\vskip 1 cm

\noindent {\large {\bf 3. Dynamic phase transitions in the Ising ferromagnet:}}

\noindent {\it (a) Model and simulation}

To investigate the true dynamic phase transitions (incorporating the fluctuations) one simple choice may be the kinetic 
Ising model in presence of sinusoidally oscillating (in time but uniform over the space) magnetic field studied 
by Monte Carlo simulation. For this
choice, one can take the following Hamiltonian,

\begin{equation}
H = -J {\Sigma_{<ij>}}S_i S_j - h(t) \Sigma_i S_i
\label{ihm}
\end{equation}

\noindent where, the first term represents the spin-spin the Ising type ferromagnetic interaction. 
$S_i (\pm 1)$ is Ising spin at $i$-th site of the lattice, $J (> 0)$ is nearest neighbour 
ferromagnetic interaction strength.  The second term represents the spin-field interaction. 
$h(t) = h_0 {\rm cos}({\omega}t)$ is applied oscillating magnetic field.
$h_0$ is the amplitude and $\omega (= 2\pi f)$ is the angular frequency of the oscillating field.
The boundary condition is periodic in all sides of the lattice.

Due to the presence of second term in the Hamiltonian, the system always remain far from equilibrium. However,
the dynamical evolution of the system can be studied by Metropolis algorithm \cite{mc}. 

The studies on Ising systems in the presence of 
an oscillating magnetic field have been made mostly by employing the Monte Carlo method  using Metropolis
single spin flip dynamics\cite{mc}. Starting from a (random or perfectly ordered) initial configuration of
spins, the spin state $S_i(t)$ at any site $i$ and in any time $t$ for a fixed temperature $T$ 
is updated (sequentially or randomly) with the
following probability function \cite{mc}
\begin{equation}
W(S_i \to -S_i) = {\rm Min}[1, {\rm exp}(-{{\Delta H} \over {K_BT}})]
\label{metro}
\end{equation}

\noindent where $\Delta H = 2S_i[\Sigma_j S_j(t) +h(t)]$, is the change in energy due to spin flip, $K_B$ is Boltzmann's 
constant and $T$ is the temperature. 
First the instantaneous response magnetisation per lattice site at time $t$ is easily calculated: 
$m(t) = \Sigma_i S_i(t)/N$, where $N$ is the total number of spins in the lattice ($N = L^d$ if one considers
a $d$-dimensional hypercubic lattice of linear size $L$). 
$N$ such spin updates is defined as one MC step per spin(MCSS). This is the unit of time in this simulational study.
After that, one can study the dynamical response of the system.
In this chapter, only the dynamic transitions, out of various kinds of dynamical responses, will be discussed. The dynamical 
order parameter (as defined in ref\cite{tom}) 
can be calculated as $Q =  \oint m(t) dt/\tau$, where $\tau = (1/f)$ ($f$ is linear frequency)
is the time period of the applied oscillating magnetic field. The magnitude of the magnetic field is measured in the
unit of $J$ and the temperature is measured in the unit of $J/K_B$. It should be mentioned here that the results are 
independent of the dynamics (Metropolis, Glauber etc.) employed to study the dynamic transitions.

\vskip 0.5 cm

\noindent {\it (b) Dynamic transition:}

The value of stabilised $Q$ is calculated \cite{ac} for fixed values of $T$, $h_0$ and $f$. It was observed that for a fixed
frequency $f$, the dynamic order parameter $Q$ is nonzero for lower values of $T$ and $h_0$ and it would vanish for
higher values of $T$ and $h_0$. For lower values of temperatures and field amplitudes the system is dynamically
ordered and it loses its dynamical ordering for higher values of $T$ and $h_0$. This dynamical order-disorder transition
is associated with the breaking of symmetries of $m-h$ loop . For lower values of $T$ and $h_0$, the magnetisation
oscillates asymmetrically (around $m=0$ line) which gives rise 
to asymmetric $m-h$ loop (resides asymmetrically in $m-h$ plane) (see Fig.\ref{dpt}).
In this case $Q \neq 0$, and this phase is a symmetry-broken phase. When $T$ and $h_0$ become large, the magnetisation
oscillates symmetrically (around $m=0$ line). This gives rise to symmetric $m-h$ loop (resides symmetrically in $m-h$ plane)
and consequently $Q=0$. This phase is called symmetric phase. In this regard, one may call that this dynamic transition
is associated with a dynamical symmetry breaking. 


Extensive Monte Carlo (MC) simulation  was performed \cite{rev} to study this dynamic transition. 
MC simulation was done in the Ising ferromagnet in two and three dimensional hypercubic lattices. The dynamic
transition was observed and the dynamic phase boundary was drawn in $h_0-T$ plane taking frequency as a parameter. 
The dynamic phase boundary in both the dimensions (two and three) are qualitatively similar in nature. Along the phase
boundary the dynamic transition occurs at higher temperatures for lower values of field amplitudes. The high 
temperature (low field) transitions are continuous and the low temperature (high field) transitions are discontinuous.
A point (marked TCP in figure \ref{ht}) separating these two types of transitions is called a tricritical point. Another important
thing should be noted in this dynamic phase boundary is, the variation of the phase boundary with respect to the frequency
of the driving field. As Fig. \ref{ht} shows, the phase boundary shrinks inward as the frequency decreases. 
As an extrapolation
of this senario, one may conclude that, in the static limit (zero frequency limit) the dynamic transition disappears. This
is the important significance of {\it true} dynamic transition.


The dynamic transition is, in fact, a manifestation of the coercivity property (one of the important features of
hysteresis)\cite{ma6,stauffer}. In the $Q \neq 0$ phase, the $m-h$ loop is not symmetric about the field axis and lies in the
upper (or lower) half of the $m-h$ plane depending on their initial magnetisation. A minimum
magnitude of external and opposite magnetic field (coercive field) is required to change the sign of the magnetisation
for complete reversal within the time period of the oscillating magnetic field. 
This magnitude of the coercive field depends on the temperature $T$. The magnitude of coercive
field increases as the temperature decreases. In the case of sinusoidally oscillating magnetic field, for a transition
to a $Q = 0$ phase from $Q \neq 0$ phase, the field amplitude should be at least of the order of coercive field depending 
upon the temperature $T$. So, in a sense, the dynamic phase boundary (in the low frequency limit) is the coercive field
variation with respect to the temperature. Since, $h(t) = h_0 {\rm cos}({\omega t})$ and $|h(t)| \leq h_0$, the phase
boundary, is the upper bound of the coercive field variation with respect to the temperature $T$. However, the difference
of this upper bound increases with increasing frequency, because of the effectice relaxational lag ($\tau_{eff}$, 
time lag of magnetisation with respect to the applied field). The effective 
relaxational lag of the magnetisation arises due to the intrinsic relaxation time of the system. In fact, the tricritical
point $T_d^{TCP}(h_0,\omega)$ on the phase boundary appears because of the system's failure to relax within the time
period $2\pi/\omega$ of the oscillating field. The intrinsic relaxation time in the ferromagnetic phase decreases with
lowering of temperature and below $T_d^{TCP}(h_0,\omega)$, $\tau_{eff} \leq 2\pi/\omega$ (equality at $T = T_d^{TCP}$).
So, that the magnetisation changes sign (from $+m$ to $-m$) abruptly and consequently $Q$ changes from a 
value very near to unity to zero discontinuously. This indicates the TCP should decrease with higher frequency.
At zero temperature, the
transition is completely mechanical (purely field driven; without any thermal fluctuation) and can only be a discontinuous
one. Above $T_d^{TCP}(h_0,\omega)$, the thermal fluctuations win-over and determine the continuous nature of the transition.
There is a controversy regarding the existence of TCP. The details of studies on TCP will be discussed in the subsection
where the relation between dynamic transition and stochastic resonance, is discussed.

It should be mentioned here that another type of dynamic phase transition was studied \cite{arnab} in the kinetic Ising model
with negative pulsed magnetic field of finite duration. But in this chapter, the dynamic phase transition, only due to 
sinusoidally oscillating magnetic field, will be discussed.
\vskip 0.5cm

\noindent {\it (c) Is dynamic transition a phase transition ?}

After getting the dynamic transition and the dynamic phase boundary, the following questions naturally arise:
{\it Is dynamic transition a phase transition ? Is there any evidence of the divergence of
time scale and length scale at the transition point ? What will be the behaviour of 'dynamic specific
heat' and fluctuations of dynamic order parameter near the transition point ?} The studies related
to the above questions will be discussed in this subsection. 

The critical slowing down is an important phenomenon observed in equilibrium transition (ferro-para)
which indicates the divergence of time scale (relaxation time) at transition point. In the MC simulation if the initial 
condition is all spins directed upward and the sinusoidal magnetic field is driving the system, the
$m-h$ loop gets stabilised after a transient behaviour. As a result, the dynamic order parameter $Q$
has also a transient behaviour. 

It has been observed carefully \cite{ma1} that the dynamic order parameter $Q$
does not acquire the stable value within the first cycle of the oscillating
field. It takes several cycles (of the oscillating field) to get
stabilised i.e., it shows 'relaxation' behaviour.
Starting from the initial (all spins are up) configuration, the $Q$ 
has been calculated for various numbers (say $n$-th) of cycles of the oscillating
magnetic field and plotted (inset of Fig.\ref{relax}) against the number of cycles ($n$). 
Each value of $Q$ shown here has been obtained by averaging over 100 random
Monte Carlo samples.  Inset of Fig.\ref{relax} shows a typical 'relaxation' 
behaviour of the dynamic order parameter $Q$. 
This has been plotted for fixed values
of $\omega = 2\pi \times 0.04$, 
$h_0$= 1.0 and  $T$= 1.5. It shows that the dynamic order
parameter $Q$ is relaxing as the time (number of cycles) goes on. 
The best fit curve shows that the 'relaxation' is exponential type.
So, one can write $Q \sim Q_0 ~~{\rm exp}(-n/\Gamma)$, where $\Gamma$ is the
'relaxation' time which provides the 'time scale' for this prototype of nonequilibrium
problem.
The physical interpretation of $\Gamma$ is,
the number of cycles required, so that $Q$ becomes $1/e$ times of
its initial value (value of $Q$ at starting cycle). 
From the exponential
fitting, the 'relaxation' time ($\Gamma$) has been measured.
The temperature ($T$) variation, for fixed
values of $\omega$ and $h_0$, of this 'relaxation' time 
$\Gamma$ has been studied
(in the disordered region of dynamic transition)
and displayed in Fig.\ref{relax}. 
The temperature ($T$) variation of $\Gamma$ are shown (Fig.\ref{relax}) for
two different values of $h_0 (= 0.5 {\rm~~and~~} 1.0)$ and for a fixed
value of $\omega = 2\pi \times 0.04$ here.
From the figure (Fig.\ref{relax}) it is clear that the relaxation time $\Gamma$ 
diverges near the dynamic transition point (where $Q$ vanishes) in the both
cases ($h_0$ = 0.5 and 1.0). This study \cite{ma1} first
indicated that this dynamic transition is associated with a diverging 'time scale'.


An analytical formulation of this critical slowing down (of $Q$) can be done \cite{ma1}
by solving the linearised mean-field equation of motion for the average magnetisation .
In the limit of $h_0 \to 0$ and $T > 1$, the equation (eqn. \ref{mf}) can be
linearised (i.e., linearising tanh term) as 
\begin{eqnarray*}
\tau {dm \over dt} = -\epsilon m + {{h_0 {\rm cos}(\omega t)} \over T},
\end{eqnarray*}
\noindent where $\epsilon = 1 - 1/T$.
The solution of the above equation is
\begin{eqnarray*}
m(t) = {\rm exp}(-\epsilon t/\tau) + m_0 {\rm cos}(\omega t -\phi),
\end{eqnarray*}
\noindent where $m_0$ and $\phi$ are two constants.
The value of the dynamic order parameter $Q$ at $n-$th cycle of the
oscillating field is
\begin{eqnarray*}
Q = {{\omega} \over {2\pi}} \oint m(t) dt = {{\omega} \over {2\pi}}
\int_{t_{n-1}}^{t_{n}} m(t) dt,
\end{eqnarray*}
\noindent where $t_n = 2\pi n/\omega$.
The value of $Q$, at the $n$-th cycle, can be written as
\begin{eqnarray*}
Q = Q_0 ~~ {\rm \exp}(-{{2\pi n \epsilon} \over {\tau \omega}}) 
= Q_0  ~~{\rm \exp} (-n/\Gamma)
\end{eqnarray*}
\noindent $Q_0$ is a constant independent of $n$. The above form shows that
$Q$ relaxes exponentially with the number of cycles ($n$) of the oscillating
field. The 'relaxation' time $\Gamma$ is equal to ${\tau \omega}/
{2 \pi} \epsilon$. 
It should be noted here that the dynamic transition occurs at $T$ = 1 in the
limit $h_0 \to 0$ . So, for $h_0 \to 0$  near the dynamic transition point 
(where the linearisation holds good) the behaviour of
relaxation time is
\begin{eqnarray*}
\Gamma \sim \epsilon^{-1} \sim \left(T - T_d(h_0 \to 0)\right)^{-1}
\end{eqnarray*}
\noindent which shows the power law (exponent is unity) divergence 
of the 'relaxation' time at
the dynamic transition point. It may be noted here that the critical slowing
down discussed above ($Q$ relaxes/vanishes exponentially) 
is valid only for $Q = 0$, not for positive $Q$.

\vskip 0.5 cm

\noindent {\it (d) Behaviour of dynamic 'specific-heat' near transition point}

The total energy averaged over a cycle of the oscillating magnetic field can be written as \cite{ma1}: 
$E_{tot} = \oint H dt/\tau$, where $H$ is the Hamiltonian (equation \ref{ihm}) of the system. 
The dynamic 'specific heat' $C_{tot}$ can be defined as the temperature derivative of the total energy
$E_{tot}$. Now, if $C_{tot} (= {{dE_{tot}}/{dT}}$) is plotted against the temperature, the plot
shows (Fig.\ref{spht}) a very sharp peak at the transition point in believe that it will diverge for infinite system. 
This behaviour is similar to that observed in well
known equilibrium phase transitions. One can detect the dynamic transition and can have an estimate
about the transition point from the temperature variation of the response like dynamic 'specific heat'.


In the equilibrium phase transition it is well known that the specific heat is related to the fluctuation
in energy. What one should expect in the case of this type of nonequilibrium transition ? To have a direct
answer to this question the fluctuation in 'energy' is studied as a function of temperature. The fluctuation in
total energy is: $\delta E_{tot}^2 =\left(<E_{tot}^2> - <E_{tot}>^2\right)$. Here the symbol $<...>$ denotes
the average over various samples (obtained from the different MC samples).
If this quantity is plotted against the temperature, it would also get sharply peaked at the transition
temperature \cite{ma2}. 


\vskip 0.5cm
\noindent {\it (e) Evidence of diverging length scale}

The evidence of dynamic correlation length of this type of dynamic transition observed in the Ising like
extended system was also reported \cite{rik2}. The dynamic susceptibility (in 2D) is defined as
$\chi = L^2 [<Q^2> - <Q>^2]$, keeping in mind that the fluctuation-dissipation theorem holds good \cite{ma2}
for this type of nonequilibrium transition also. This is plotted against $1/R = (2\pi/\omega)/<\tau(h_0)>$
where $<\tau(h_0)>$ is average lifetime or nucleation time of the system. $\chi$ was plotted against $1/R$
for various values of $L (=64, 90, 128)$. The figure shows the peak of $\chi$ increases in 
height with increasing system size ($L$). This clearly indicates the finite size effects in $\chi$
and implies the existence of a divergent length scale associated with the order parameter correlation function
near the transition point. It is important to note here that this study was done by varying $1/R$ (keeping $T$
and $h_0$ fixed) whereas most of the studies on dynamic transition have been done by varying the temperature
$T$ (keeping $\omega$ and $h_0$ fixed). However, the results are qualitatively invariant under the choice of
tunable parameter. This study \cite{rik2} gives an important idea regarding the divergence of 'length scale'
at the transition point of this dynamic transition.

\vskip 0.5 cm

\noindent {\it (f) Dynamic phase transition and hysteresis}

\noindent {\it (i) Analytic forms of
the loop area and the dynamic correlation near
the transition point}

\noindent In the earlier sections it was discussed that the dynamic transition has a very close resemblance
with the magnetic hysteresis. In this section, the relations among the hysteretic loss, the dynamic order 
parameter and the dynamic correlations are dicussed \cite{ma3}. 

\noindent The form of the oscillating magnetic field is
\begin{equation}
h(t) = h_0 \cos(\omega t).
\end{equation}

\noindent The dynamic order parameter is defined as
\begin{equation}
Q = {\omega \over 2\pi} \oint m(t) dt,
\end{equation}
\noindent which is nothing but the time-averaged magnetisation over a
full cycle of the oscillating magnetic field.
The hysteresis loop area is
\begin{equation}
A = -\oint m dh
 = h_0 \omega \oint m(t) \sin(\omega t) dt,
\end{equation}
\noindent which corresponds to the energy loss due to the hysteresis.
The Dynamic correlation is defined as
\begin{eqnarray}
C_d = <m(t)h(t)> - <m(t)><h(t)>, \nonumber
\end{eqnarray}

\noindent where $<..>$ denotes the time average over the full cycle of
the oscillating magnetic field. Since $<h(t)>$ = 0, one can write

\begin{equation}
C_d = {\omega \over 2\pi} \oint m(t) h(t) dt 
= {{\omega h_0} \over {2\pi}} \oint m(t) \cos(\omega t) dt.
\end{equation}
\noindent The dynamic correlation has another physical interpretation. For the
cooperatively interacting spin system, this is the negative of the time
averaged spin-field interaction energy (per spin) in d-dimensions ($<E_f> = -{{\omega} 
\over {2\pi L^d}}\oint 
\sum_i \sigma_i ~h(t) dt$) over a complete
cycle of the oscillating field.

In the dynamically disordered ($Q = 0$)
phase and near the transition point, 
the time series of the magnetisation ($m(t)$) can be approximated
as a square wave with a phase lag $\delta$ with the applied sinusoidal
magnetic field. This approximation works well in the low temperature region.

\begin{equation}
m(t)=
\left\{ 
\begin{array}{rcl}
 1 & ~~{\rm for}~~& 0 < t < \tau/4 + \delta/\omega  \\
-  1 &  ~~{\rm for}~~ &\tau/4+\delta/\omega 
< t < 3\tau/4+\delta/\omega \\
  1 & ~~{\rm for}~~&3\tau/4+\delta/\omega < t < 2\pi/\omega,
\end{array}\right.
\label{dis}
\end{equation}

\noindent where $\tau$ is the time period of 
the oscillating field and $\delta$ is the phase
lag between magnetisation $m(t)$ and the magnetic field $h(t)= h_0 \cos (\omega
t)$. The value of the hysteresis loop area can easily be calculated as
\begin{equation}
A = 4 h_0 \sin(\delta).
\end{equation}
\noindent This form of the loop area
was also obtained  from the assumption that it is approximately
equal to four times the product of coercive field and remanent
magnetization
(here the remanent magnetisation equal to unity),
where the coercive field is identified as $h_0 \sin(\delta)$
(the change in field during the phase lag). Considering the 
same form of the magnetisation the dynamic correlation $C$ can also be
calculated exactly as 
\begin{equation}
C_d = {{2h_0} \over {\pi}} \cos(\delta).
\end{equation}

\noindent From the above forms of $A$ and $C_d$ it can be written as
\begin{equation}
{{A^2} \over {(4h_0)^2}} + {{C_d^2} \over {(2h_0/\pi)^2}} = 1.
\end{equation}
\noindent The above relation shows that the loop area 
$A$ and the dynamic correlation
$C_d$ is elliptically related to each other. 

The qualitative nature of the dynamic phase boundarycan be realised by considering the simplified
form of the instantaneous magnetisation in the ordered phase.
The dynamically ordered region ($Q \neq 0$) can be approximated by considering the
following form of the magnetization

\begin{equation}
m(t)=
\left\{ 
\begin{array}{rcl}
 1 & ~~{\rm for}~~& 0 < t < \tau/4 + \delta/\omega  \\
 1-m_r &  ~~{\rm for}~~ &\tau/4+\delta/\omega 
< t < 3\tau/4+\delta/\omega \\
  1 & ~~{\rm for}~~&3\tau/4+\delta/\omega < t < 2\pi/\omega.
\end{array}\right.
\label{ord}
\end{equation}
\noindent In the above simplified approximation, 
it was considered (since $Q \neq 0$) that the magnetisation
can not jump to the other well, however the value of initial magnetisation is
reduced by the amount $m_r$. 
In the real situation it has been observed that this well is not fully
square (as assumed above in the form of $m(t)$), it has a cusp like 
(or parabolic) shape.
For $m_r$ = 2, the above functional form of $m(t)$ will take the form of
eqn. \ref{dis} and one can get the disordered ($Q=0$)
phase. Taking the above form of magnetisation the dynamic order parameter
$Q$ can be calculated as $Q = (2 - m_r)/2$. It may be noted that, in this
simplified approximation the dynamic order parameter $Q$ is independent of
phase lag $\delta$, which is not observed in the real situation (phase lag
shows a peak at the transition point) \cite{ma3}. However, this simple picture can 
anticipate the convex (looking from the origin) 
nature  of the dynamic phase boundary. As the
temperature increases $m_r$ increases and it also increases as the field
amplitude increases. 
Since $m_r$ increases as $h_0$ and $T$ increases,
in the simplest linearised assumption, one can consider $m_r$ is
proportional to the product of $h_0$ and $T$. Demanding, $m_r = 2$ for
the dynamic transition ($Q = 0$), one can readily obtain $(h_0)_d T_d$ =
constant. This equation tells that
the dynamic phase boundary will be convex having the shape of rectangular hyperbola ($xy=c$). 
The convex  
nature of the phase boundary remains invariant even
if one
assumes that $m_r$ is any increasing function of 
both $T$ and $h_0$ (for example,
power law; $m_r \sim T^x h_0^y$, in this particular case the equation
of the dynamic phase boundary becomes $T_d^x (h_0)_d^y$ = constant, it
is easy to see that this gives the convex shape
of the dynamic phase boundary). 
However, this very
simple assumption can not describe the entire form of the phase boundary 
accurately, particularly near the end points ($(h_0)_d$ = 0 
and $T_d$ = 0).

\vskip 0.5 cm

\noindent {\it (ii) General relation among Dynamic order parameter, Hysteresis loop area
and the Dynamic correlation}

From the usual definitions (given in earlier section) 
of $C_d$ and $A$, one can write
\begin{eqnarray}
{1 \over {{\sqrt {2\pi}}}}
\left({{2\pi C_d} \over {\omega h_0}} - i{{A} \over {\omega h_0}}\right) = 
{1 \over {{\sqrt {2\pi}}}}
\oint m(t) \exp
({-i \omega t}) dt, \nonumber
\end{eqnarray}
\noindent where $m(\omega) = {1 \over {\sqrt {2\pi}}} 
\oint m(t) \exp({-i \omega t}) dt$.
\noindent
So,
\begin{flushleft}
$C_d = {{h_0 \omega} \over {\sqrt {2 \pi}}}{\rm Re}\left(m(\omega)\right)$
\end{flushleft}
\noindent and
\begin{flushleft}
$A = -{h_0 \omega {\sqrt {2\pi}}}{\rm Im}\left(m(\omega)\right)$.
\end{flushleft}

\noindent The general (complex) form of $m(\omega ')$ will be
\begin{flushleft}
$m(\omega') = |m(\omega')| \exp (i\phi)$
\end{flushleft}
\begin{flushleft}
$m(\omega') = {1 \over {\sqrt {2\pi}}}
\left({{4\pi^2 C_d^2} \over {h_o^2 \omega'^2}} + {{A^2} \over {h_0^2
\omega'^2}}\right)^{1/2} \exp i\left[-\tan^{-1} {A \over {2\pi C_d}}\right]$
\end{flushleft}

\noindent Note that the phase $\phi$ of $m(\omega')$ is independent of $h_0$ and $\omega$. 
So, $Q$ is related with $A$ and $C_d$ as follows
\begin{eqnarray*}
Q = {1 \over {\tau}}\oint m(t) dt   
\end{eqnarray*}
\begin{eqnarray*}
 ={1 \over {\sqrt {2\pi} \tau}}\int d\omega'
\oint m(\omega') \exp(i\omega' t) 
dt
\end{eqnarray*}
\begin{equation}
 ={1 \over {2\pi \tau}} \int d\omega' \oint 
{\sqrt {\left({{4\pi^2 C_d^2} \over {h_o^2 \omega'^2}} 
 + {{A^2} \over {h_0^2 \omega'^2}}\right)}} 
 {{\bf e}}^{ \left[i(\omega't - 
\tan^{-1} {A \over {2\pi C_d}})\right]} dt.
\label{gen}
\end{equation}
\noindent Above equation gives the general relationship among $Q$, $A$ and $C_d$. 

It has been observed that the steady response $m(t)$, to a sinusoidally 
oscillating magnetic field ($h(t) = h_0 \cos (\omega t)$), is
periodic (with phase lag $\delta$) and has the same
periodicity ($\tau = 2\pi/\omega$) of the field. So, one can
write $m(t)$ in a Fourier series as
\begin{equation}
m(t)= a_0 + \sum_{n=1}^{\infty} a_n \cos(n\omega t) + \sum_{n=1}^{\infty}
 b_n \sin(n\omega t).
\end{equation}
\noindent From the usual definitions of $Q$, $A$ and $C_d$, it is easy to see
that 
\begin{flushleft}
$a_0 = Q$,
$a_1 = 2C_d/h_0$ ~~{\rm and}~~
$b_1 = A/(\pi h_0).$
\end{flushleft}
\noindent So, one can write
\begin{equation}
m(t) = Q + {{2C_d} \over {h_0}} \cos (\omega t) + .....+{{A} \over {\pi h_0}}
\sin(\omega t) + ....  .
\label{mt0}
\end{equation}
\noindent Keeping only the first 
harmonic terms (ignoring higher harmonics) one can
easily express the instantaneous magnetization as
\begin{equation}
m(t) = Q + m_0 \cos(\omega t - \delta)
\label{mt}
\end{equation}
\noindent where the amplitude of magnetization is $m_0 = [(2C_d/h_0)^2 + 
(A/(\pi h_0))^2]^{1/2}$ and the phase lag is $\delta = \tan^{-1}(A/(2\pi C_d))$.

\vskip 0.5cm

\noindent {\it (g) Dynamic phase transition and stochastic resonance}

To study the relationship between dynamic transition and the stochastic resonance \cite{sr} extensive MC 
simulations were performed \cite{ma5} in the kinetic Ising model in presence of sinusoidally oscillating magnetic field.
The frequency is $f = 0.001$ (kept fixed throughout the study). So, one complete cycle of the oscillating
field takes $\tau$ = 1000 MCSS. 
A time series of magnetization $m(t)$ has been generated up to $10^6$ MCSS. This time series contains
$10^3$ (since $\tau$ = 1000 MCSS) 
cycles of the oscillating field. The dynamic order parameter $Q$ has been calculated
for each such cycle.
So, the statistics 
(distribution of $Q$) is based on $N_s = 10^3$ different
values of $Q$. 

The statistical distribution $P(Q)$ of dynamic order parameter $Q$
and its temperature dependence
have been studied \cite{ma5} closed to the phase boundary to detect the nature\cite{binder} 
of the transition.
Fig.\ref{qdst}a shows the distributions $P(Q)$ (at fixed
value of the field amplitude) for three different values of temperature. Below, the transition 
the distribution shows only two equivalent peaks
centered around $\pm 1$. Close to the transition point, a third peak
centered around zero is developed. As the temperature increases slightly, the strength of the
third peak increases in cost of that of two other (equivalent) peaks.
Above the transition, only one peak is observed centered around zero. 
This indicates  that the
transition is first order or discontinuous.

What is the origin of this kind of first order transition ? To get the answer of this question,
the time variation of the magnetization $m(t)$ is studied \cite{ma5} for several cycles of the oscillating
magnetic field $h(t)$, close to the transition. 
Sometimes, the system likes to 
stay in the positive well (of the Landau type double well form of the free energy) 
and sometimes it likes to stay in other. It
is obvious that the best time for the system to switch from one well to the other one, is
when the value of the field is optimum ("good opportunity") .
So, if the system misses one "good opportunity" (first half period of the oscillating field)
to jump to the other well it has to wait for a new chance (another full period of the oscillating field).
Consequently, it shows that the residence time (staying time in a particular well) can only be 
nearly equal to an odd
integer multiple of the half-period (half of the time period of the oscillating field).
This leads to two consequences: 

(1) The distribution of the 
dynamic order parameter $Q$ would be peaked around three values (i) $Q \approx 0$, when the system 
utilizes "good opportunity" and goes from one well to the other (marked 'A' in
Fig.\ref{qdst}a), (ii) $Q \approx -1$, when the
system misses the "good opportunity" to go from the negative well to the 
positive well and it stays for one (or
more) full period in the negative well (marked 'B' in Fig.\ref{qdst}a), 
(iii) $Q \approx +1$, when the system misses the "good opportunity" to
go from the positive well to the negative well and spends one (or more) full period
in the positive well (marked 'C' in Fig.\ref{qdst}a). 
As a result, the distribution of $Q$ would give three distinct peaks centered
at +1, -1 and 0.

(2) The other consequence of this kind of time variation, of magnetization $m(t)$, is
the "stochastic resonance" \cite{sr}. 
This can be detected from the distribution of residence time (the time system spends in a particular
well).
The distribution ($P_r$) of residence time ($\tau_r$) 
will be peaked multiply around the odd integer multiple of
half-period.
One such distribution is shown in Fig.\ref{sr}. The distribution shows multiple peaks around the
odd integer values (500, 1500, 2500, 3500, 4500 and 5500 MCSS)
of half-period ($\tau/2$=500 MCSS, of the driving fields). The heights of the peaks decreases
exponentially (dotted line in Fig.\ref{sr} ) with the peak positions. This is the fingerprint
of stochastic resonance \cite{sr}.

Figure \ref{qdst} shows the distributions of the dynamic order parameter $Q$
for three different values of the temperature. Here, the field amplitude $h_0$ is quite low in
comparison with that used
in the earlier case (Fig.\ref{qdst}a).
It shows that, in the ordered region, this gives two (equivalent)
peaks (Fig.\ref{qdst}b) and as the temperature increases these two peaks come close to each other continuously 
(Fig.\ref{qdst}b) and close 
to the transition (and also above it) (Fig.\ref{qdst}b)
only one peak (centered around zero) is observed. This feature reveals the continuous
or second order transition. Hence, it was proved that a tricritical point 
would exist which separates the nature (discontinuous/continuous) of the dynamic
transition. However,  a recent study \cite{rik3} claims that the existence of TCP \cite{ma5}
is a correct observation but a finite size effect. 
For small system size the distribution of dynamic order parameter $Q$ shows three peaks very 
close to the transition point. This was observed earlier \cite{ma5} for small system size. 
However,  
the distribution of $Q$ has only two peaks near the transition
point for much larger systems revealing only the continuous nature of the transition.

\vskip 0.5cm

\noindent {\it (h) Dynamic phase transition for randomly varying field:}

 Very recently, an interesting version of this dynamic phase
transition has been predicted \cite{ma4} in a ferromagnetic Ising system
when the external field on the system varies in time stochastically.
The long time
response (magnetisation) of a kinetic Ising system represented by the
Hamiltonian (eqn. \ref{ihm}) is studied when the uniform field over the sample $h(t)$ 
varies randomly in time with a uniform distribution bounded between
$+h_0/2$ and $-h_0/2$. In a Monte Carlo simulation study in two
dimension, the nature of the response magnetisation
(see Fig.\ref{rnd}a and Fig.\ref{rnd}b) is studied with 
the dynamic order parameter $Q (= (1/\tau)\int 
_0^{\tau} m(t') dt'; ~ \tau
\gg 1)$ which is given by the long-time average (over the active duration
of the magnetic field) of magnetisation.
It was found that $Q$ assumes nonzero values below a phase boundary
line in the $h_0 - T$ plane, and vanishes continuously at the
transition boundary (see Fig.\ref{rnd}c). Again, the dynamic symmetry
breaking transition occurs due to the competing time scales; the
relaxation time of the many-body system being larger than the switching
time of the random field. Such a dynamic 
transition is again a nonequilibrium
transition, very similar to that for oscillating fields discussed
earlier.  It may be mentioned
that, in  a slightly different context, a discrete map
version of the mean field equation of motion (eqn. \ref{mf}) with similar
stochastically varying field $h(t)$ was analysed recently by
Hausmann and Ruj$\acute {\rm a}$n \cite{haus}. The dynamic transition for a randomly
varying magnetic field was also studied \cite{ma4} by solving the mean-field equation
(eqn. \ref{mf}) of motion of average magnetisation. 

\vskip 0.5 cm

\noindent {\it (i) Athermal dynamic transition in random field Ising model:}

The kind of nonequilibrium dynamic transition discussed so far was assisted by thermal 
fluctuation. An interesting phenomenon, the athermal hysteresis, has been studied \cite{shukla}
recently in
random field Ising model. Now one may ask, is there any disorder induced dynamic transition observed at $T=0$ ? To
investigate this the random field Ising model (in 2D) in presence of oscillating magnetic
field was studied \cite{rfim} at $T=0$ by MC simulation.
A square lattice of linear size $L$ is taken. Each site is labelled by
an integer $i$ and carries an Ising spin $S_i$ ($S_i = \pm1$) which interacts
with all its nearest neighbours (spins) with a ferromagnetic interaction
strength $J$. At each site $i$, there is a local {\it quenched} random field
$h_i$. The random fields 
${h_i}$ are assumed to be independent and identically distributed
random variables with a rectangular probability distribution $P(h_i)$. The
random field
$h_i$ can take any value from $-w/2$ to $+w/2$ with the same probability. The
width of the distribution is $w$.
In addition, there is a uniform (in space) magnetic field $h(t)$ which is
varying sinusoidally ($h(t) = h_0 \cos(\omega t)$) in time. The amplitude and
the frequency are denoted by $h_0$ and $\omega$ respectively. This kind of
model is described by the Hamiltonian
\begin{equation}
H = -J \sum_{<ij>} S_i S_j - \sum_i h_i S_i - h(t) \sum_i S_i,
\end{equation}
\noindent under the periodic boundary condition. 

The {\it zero-temperature} single spin-flip dynamics is specified by the
transition rates ($W$)
\begin{equation}
W(S_i \to -S_i) = \Gamma, ~~{\rm if~~} \Delta E \leq 0 {~~~\rm and~~}
W(S_i \to -S_i) = 0, {~~\rm otherwise}
\end{equation}
\noindent where $\Delta E$ is the change in energy due to spin flip. In 
words the algorithm is: never flip the chosen spin  
if this process would increase the energy and flip otherwise. We have
started with all spins are up ($S_i = +1$) as an initial condition and updated
the lattice sequentially using the above flipping algorithm. One such full
scan over the entire lattice consists a Monte Carlo 
step per spin (or MCS). The
instantaneous magnetisation (per site) $m(t)$ is easily calculated,
\begin{equation}
m(t) = {1 \over {L^2}} \sum_i S_i.
\end{equation}
\noindent After an initial transient period the intantaneous magnetisation
$m(t)$ has been found to be stabilised and 
periodic with the same periodicity of the applied
oscillating field. 
For a particular values of $h_0$, $\omega$ and $w$ the 
dynamic order parameter $Q (= {1 \over {\tau}}\oint m(t) dt)$ is calculated by averaging over 20 different
random disorder (quenched) realisations.

The simulations are performed on a square lattice of linear size
$L = 100$ and a particular value of the frequency 
($\omega = 0.01 \times 2\pi$) of the oscillating
magnetic field. The time is measured in units of Monte Carlo steps per
spin or MCS and the values of random field and the oscillating field are
measured in units of interaction strength $J$.

It has been observed numerically that,
for fixed values of $h_0$ and $w$, 
the magnetisation becomes periodic (in time) with the same periodicity as
the applied sinusoidal magnetic field. For the smaller values of the quenched
disorder ($w = 8.0$) and the field amplitude ($h_0 = 0.5$), the magnetisation
oscillates asymmetrically about the zero line 
i.e., the system remains in a dynamically symmetry-broken
phase. Consequently, the
hysteresis ($m-h$) loop 
resides on the upper half plane formed by $h(t)$ and
$m(t)$. So, the time-averaged magnetisation over a full cycle of
the oscillating field, the dynamic order parameter, is nonzero
in the symmetry-broken phase. By increasing 
the field amplitude (for $h_0 = 2.0$) 
keeping $w$ fixed ($w = 8.0$), it was observed that the system 
acquires a 
dynamically symmetric phase, i.e., the magnetisation oscillates 
symmetrically about the zero line.
The hysteresis loop is also symmetric. Consequently, the
value of the dynamic order paramater $Q$ is zero in this dynamically
disordered (symmetric) phase.

In the dynamically disordered phase, the dynamic order parameter
$Q$ can be 
kept at zero in two ways, either by increasing the random field width $w$
for a fixed field amplitude $h_0$ or vice versa. 
So, in the plane formed by
the field amplitude ($h_0$) and the width ($w$) of the 
quenched disorder (random field), one can think of a
boundary line, below which $Q$ is nonzero and above
which it vanishes. Figure \ref{dpt0}(a) displays such a phase boundary in the $h_0-w$
plane obtained by Monte Carlo simulation. 
The nature (discontinuous/continuous) of the transition depends on the
value of $w$ and $h_0$ on the phase boundary line. The transition across
the upper part of phase boundary line is discontinuous and it is continuous
for the rest part of the boundary. A tricritical point 
 on the phase boundary
line separates these natures. Figure \ref{dpt0}(b) demonstrates two typical transitions
for two sets of values of $w$ and $h_0$ lying 
just in the left and right sides of the tricritical
point (TCP). In the case of discontinuous transition, the dynamic order
parameter $Q$ jumps to a small nonzero value and then vanishes continuously.
The uncertainty in the location of the TCP on the
phase boundary are shown by the circle enclosing it.
It was not yet checked whether this TCP observed here is a finite size effect or not.
\vskip 1 cm

\noindent {\large {\bf 4. Dynamic phase transitions in the Heisenberg ferromagnet:}}

\noindent {\it (i) Why Heisenberg model ?}

When the nonequilibrium dynamic transition is studied in the Ising model, it has some limitations.
The Ising model is a special case of general magnetic model \cite{book}, for example, the Heisenberg
model. The Heisenberg model (with ferromagnetic interactions) having uniaxial anisotropy has some
general properties which cannot be found in the Ising model. But in the limit of infinite anisotropy,
the Heisenberg model can be mapped into the Ising model. So, the natural expectation is, the Heisenberg
model with uniaxial anisotropy can be studied to have the detailed and general microscopic view
and the results can be checked in the limit of infinite anisotropy (which will give the results
in the Ising model). In this case of dynamic transitions, mainly in the magnetic model system in presence
of a magnetic field, oscillates sinusoidally in time, the Heisenberg model can serve a better role
than an Ising model. 
It would be quite interesting to know the dynamic response of uniaxially anisotropic Heisenberg
model in presence of a magnetic field, applied in different directions.
On the other hand, there is another advantage. The results obtained in the Ising
model are well established \cite{rev}. These results can be used to check the results obtained in the Heisenberg
model by approaching the limit of infinite anisotropy. This prompted to study the dynamic transition in the Heisenberg
model with uniaxial and single-site anisotropy. Recently, the dynamic transition was studied \cite{ijmpc} in the uniaxially
anisotropic ferromagnetic Heisenberg model and it was observed that the dynamic
symmetry of the order parameter component (along the anisotropy direction) can be broken in presence
of a magnetic field, applied along the direction which is perpendicular to the direction of anisotropy. 
This transition was named {\it off-axial} transition.
The transition 
is found to be continuous and the transition temperature increases as the strength of anisotropy increases. 

So,
the questions naturally arise what would be the difference in the dynamic transitions in presence of
a field, applied only along the direction of anisotropy ? How the symmetry breaking takes place ? What
would be the nature (continuous or discontinuous) of the transition ? More interestingly, what
would happen in the infinitely anisotropic case and in the Ising case ? 
To get the answers of these questions, the researchers studied the dynamic phase transition in classical
vector spin models. The dynamical phase transition in anisotropic XY-ferromagnet in an oscillating magnetic field 
is studied recently \cite{yasui} by solving the time-dependent Ginzburg-Landau equation. Very recently, it
was observed theoretically \cite{epl} that the symmetry of the vector spin model can be tailored by applying
oscillating magnetic field. For example \cite{epl}, depending on the frequency and amplitude of the field the Heisenberg
ferromagnet can behave like XY-ferromagnet.
The dynamic transitions in presence of the axial field 
(i.e., the magnetic field is applied only along the direction of anisotropy) and the off-axial field
(i.e., the magnetic field is applied only along the direction which is perpendicular to the direction
of anisotropy) are studied
\cite{ijmpc} by Monte Carlo simulation using Metropolis rate. Also, a comparison between axial
and off-axial transitions has been made and the results (in the limit of infinite anisotropy)
for both cases are compared with that observed in
the Ising model. By the application of polarised magnetic field the multiple dynamic transitions
were observed \cite{ma7} in anisotropic ferromagnetic Heisenberg model.

\noindent {\it {(ii). The description of the model}}

\vskip 0.5cm

\noindent The Hamiltonian of a classical anisotropic (uniaxial and single-site) Heisenberg model 
with nearest neighbour ferromagnetic interaction
in the presence of a magnetic field can be written as

\begin{equation}
H = -J \sum_{<ij>} \vec S_i \cdot \vec S_j -D \sum_{i} (S_{iz})^2
-{\vec h} \cdot {\sum_{i} {\vec S_i}},
\end{equation}
\noindent where ${\vec S_i} [S_{ix},S_{iy},S_{iz}]$ represents a classical spin vector of
magnitude unity situated at the $i$-th lattice site. So, $S_{ix}^2+S_{iy}^2+S_{iz}^2 = 1$ is
an equation of a unit sphere.
Classical spin
means, this spin vector can be oriented in any direction in the
vector spin space. $J (> 0)$ is the uniform nearest neighbour 
strength of the ferromagnetic interaction. 
The factor $D$ in the second term is the strength of uniaxial ($z$ here)
anisotropy favouring the spin to be aligned along the $z$-axis. The
last term is the spin-field interaction term, where ${\vec h} [h_x,h_y,h_z]$ is the
externally applied magnetic field (uniform over the space). 
When the magnetic field is applied only along the $\alpha$ - direction,
the magnetic field component $h_{\alpha}$ (may be any one of $x$, $y$ and $z$) is oscillating
sinusoidally
in time and can be written as $h_{\alpha} (t) = h^0_{\alpha} {\rm cos} (\omega t)$,
where $h^0_{\alpha}$ and $\omega$ are the amplitude and angular frequency
($\omega = 2 \pi f$; $f$ is frequency) of the oscillating field
respectively. 
Magnetic field $|\vec h|$ and strength of anisotropy $D$
are measured in units of $J$.
The model is defined in a simple cubic lattice of
linear size $L$ with periodic boundary conditions applied in all the three directions.

\vskip 0.5cm

\noindent {\it {(iii) The Simulation technique}}

\vskip 0.5cm

\noindent The model, described above, has been studied extensively by Monte Carlo
simulation using the following algorithm \cite{uli}. 
Initial configuration is a  random spin configuration. 
Here, the
algorithm used, can be described as follows.
Two different random numbers $r_1$ and $r_2$ 
(uniformly distributed
between -1 and 1) are chosen in such a way that $R^2=(r_1^2+r_2^2)$ becomes less than or equal to unity.
The set of values of $r_1$ and $r_2$, for which $R^2 > 1$, are rejected.
Now, $u={\sqrt{1-R^2}}$.
Then, $S_{ix}=2ur_1$, $S_{iy}=2ur_2$ and $S_{iz}=1-2R^2$.

Starting from an initial random spin configuration (corresponding to high temperature configuration)
the system is slowly cooled down.  
At any fixed
temperature $T$ (measured in the unit of $J/K_B$) and field amplitude $h^0_{\alpha}$ (measured in the
unit of $J$) a lattice site $i$ has
been chosen randomly (random updating). The value of the spin vector at this randomly chosen site is ${\vec S_i}$ (say). 
The energy of the system is given by the Hamiltonian (equation 1) given above. Now, a test spin vector
${\vec S_i^{\prime}}$ is chosen randomly (described by the algorithm above). For this choice of spin
vector at site $i$ the energy will be
$H^{\prime} = -J \sum_{<ij>} \vec S_i^{\prime} \cdot \vec S_j -D \sum_{i} (S^{\prime}_{iz})^2
-{\vec h} \cdot {\sum_{i} {\vec S_i^{\prime}}}$. The change in energy, associated with this change
in direction of spin vector from ${\vec S_i}$ to ${\vec S^{\prime}_i}$, is $\Delta H = H^{\prime} - H$.
Now, the Monte Carlo method \cite{book,binder} will 
decide how far this change is acceptable. The probability of the change
is given by Metropolis rate \cite{book,binder} (used here)
$W({\vec S_i} \rightarrow {\vec S^{\prime}_i}) = 
{\rm Min} [ 1, {\rm exp}(-\Delta
H/K_BT)]$. This probability will be compared with a random number $R_p$ (say)
between 0 and 1. If $R_p$ does not
exceed $W$, the move (the change ${\vec S_i} \rightarrow {\vec S^{\prime}_i}$) is accepted. 
In this way the spin vector ${\vec S_i}$ is updated.
$L^3$ such random updates of spins, defines one Monte Carlo step per site
(MCSS) and this is considered as the unit of time in this simulation. 
The linear frequency ($f = \omega/{2\pi}$) of the oscillating field
is taken 0.001 and was kept constant throughout this simulational study. 
So, 1000 MCSS is required to get
one complete cycle of the oscillating field and consequently
1000 MCSS becomes the time
period ($\tau$) of the applied oscillating magnetic field. 
To calculate any macroscopic quantity, like instantaneous magnetisation
components, the following method was employed.
Starting from an initially random configuration (which corresponds
to a high temperature phase) the system is allowed to be 
stabilised (dynamically)
up to $4 \times 10^4$ MCSS ( i.e., 40 complete cycles of the oscillating 
field) and the averages of various physical quantities are calculated
from further $4 \times 10^4$ MCSS (i.e., averaged over further 40
cycles of the oscillating field). This is quite important to get 
stable hysteresis loop and it is checked that the number of MCSS mentioned above
is sufficient to get a stable dynamic phase.
Here the total length of this
simulation for one fixed temperature $T$ is $8 \times 10^4$ MCSS
(which produces 80 complete cycles of the oscillating field). 
Then
the system is slowly 
cooled down (the value of the temperature $T$ has been reduced by 
small interval) to get the values of the statistical quantities in the low
temperature ordered phase. Here, the last spin configuration 
obtained at the previous temperature is used as the initial configuration
for the new temperature. 
The linear size ($L$) of the system has been taken equal to 20.
The CPU time required for $8 \times 10^4$ MCSS
is approximately 25 minutes on an Intel-Pentium-III processor.

\vskip 0.5 cm

\noindent {\it (iv) Off-axial dynamic transition}

\vskip 0.5 cm

The instantaneous magnetisation components (per lattice site)
$m_x ={\sum_{i} S^x_i}/{L^3}, m_y = {\sum_{i} S^y_i}/{L^3}, 
m_z = {\sum_{i} S^z_i}/{L^3}$ are calculated at each time in presence of
magnetic field.
The time-averaged (over a full cycle of the oscillating field) 
magnetisation components (the dynamic order parameter components)  
$Q_x = {1 \over \tau} \oint m_x dt, Q_y = {1 \over 
\tau} \oint m_y dt$ and $Q_z = {1 \over \tau} \oint m_z dt$

are calculated by integrating (over the complete cycle of the oscillating field) 
the instantaneous magnetisation components.
The total (vector) dynamic order parameter is expressed as
$\vec Q = {\hat x}Q_x + {\hat y}Q_y + {\hat z}Q_z.$

In this paper, two kinds of dynamic transitions were studied and compared. The {\it axial}
transition means the dynamic order parameter component $Q_z$ becomes zero from
a nonzero value at a finite temperature (the transition temperature) in presence
of a magnetic field $\vec h [0,0,h_z]$ applied only along the direction which is {\it parallel} to the direction
of anisotropy. Since the uniaxial anisotropy has been taken along the z-direction, in this case, 
the direction of magnetic field has only nonzero $z$-component.
The {\it off-axial} transition \cite{ijmpc} is the transition in presence of
a magnetic field $\vec h [h_x,0,0]$ applied only along the direction which is 
{\it perpendicular} to the direction of anisotropy. In this case,
the direction of the magnetic field has only nonzero $x$-component. 

In the case of {\it axial} transition, the instantaneous magnetisation components are
calculated at any fixed temperature $T$, strength of anisotropy $D$ and amplitude of axial
magnetic field $h_z^0$. The time-eliminated plot of $m_z-h_z$ gives the axial hysteresis
loop. It was observed that at high temperature ($T = 2.2$) the axial hysteresis loop $m_z-h_z$ is
symmetric (symmetric means the loop is distributed about $h_z$ axis in such a way that the
total $z$-component of magnetization, over a complete cycle of field, vanishes) (fig.\ref{hdt11}a). As
a result $Q_z = 0$.  And
at low temperature ($T = 1.0$) the $m_z-h_z$ loop becomes asymmetric ($Q_z \neq 0$) (fig.\ref{hdt11}b). In both cases,
the $m_x-h_z$ and $m_y-h_z$ loops lie almost along $h_z$ axis, resulting $Q_x$ and $Q_y$
equal to zero respectively. Thus a dynamic transition occurs (as the temperature decreases) at
a certain temperature from a symmetric ($Q_z = 0$; $\vec Q = 0$) to an asymmetric ($Q_z \neq 0$;
$\vec Q \neq 0$) dynamic phase in presence of an {\it axial} magnetic field. 

What was observed in the case of {\it off-axial} transition ? Recently studied \cite{ijmpc} off-axial
transition shows similar dynamic transition via breaking the symmetry of $m_z-h_x$ loop in presence of
an off-axial field (along perpendicular to the anisotropy direction i.e., x-direction). Here, at high
temperature ($T=1.8$) the $m_z-h_x$ loop is symmetric (and $Q_z=0$) and $m_x-h_x$ loop is also
symmetric ($Q_x=0$) (fig.\ref{hdt11}c). At some lower temperature ($T=0.6$), the $m_z-h_x$ loop becomes asymmetric
($Q_z = 0$) and $m_x-h_x$ loop remains still symmetric ($Q_x = 0$) (fig.\ref{hdt11}d). In both temperatures
$Q_y = 0$. So, here also a dynamic transition occurs (as the temperature decreases) at a certain temperature
from a symmetric ($Q_z = 0$; $\vec Q = 0$) to an asymmetric ($Q_z \neq 0$; $\vec Q \neq 0$) dynamic phase
in presence of an {\it off-axial} magnetic field. Interestingly, it may be noted here that in higher
temperature the $m_z-h_x$ loop is 'marginally symmetric' (lies very close to $h_x$ axis) 
rather than a symmetric loop (symmetrically distributed away from and about $h_x$ line). Strictly speaking,
the dynamic transition occurs here from a 'marginally symmetric' (loop does not widen up)
to an asymmetric phase. One can differenciate the symmetric phase from the `marginally symmetric' phase
by considering the loop area of that loop whose symmetry breaking is considered in the transition. 
In the symmetric phase, loop is sufficiently widened up resulting in a nonzero
loop area. In Fig.\ref{hdt11}a, the $m_z-h_z$ loop area is 0.686 (symmetric loop; $Q_z = 0$). 
But the `marginally symmetric' loops ($m_z-h_x$) have vanishingly small area (0.01)(see Fig.\ref{hdt11}c) and 
$Q_z = 0$. It may be noted that,
in the case of off-axial transition, if the magnetic field is applied along the x-direction only (oscillates sinusoidally
in time) the $m_x-h_x$ loop is always symmetric (consequently $Q_x = 0$) irrespective of the value of temperature and the strength of
anisotropy $D$ (z-axis). Similarly, for any field, applied along y-direction only, the $m_y-h_y$ loop is found to be
always symmetric (i.e., $Q_y = 0$) irrespective of value of $T$ and $D$. But in both cases, whether the off-axial loops
i.e., $m_z-h_x$ or $m_z-h_y$ will be symmetric (rather `marginally symmetric') or asymmetric that
 depends upon the values of temperature $T$, anisotropy $D$
and the magnetic field amplitude $h_x^0$ (or $h_y^0)$. 
These results signify that without anisotropy the dynamic transition (associated with the dynamic symmetry breaking)
cannot be observed in the classical Heisenberg model.

To investigate the dependence of transition temperature on the strength of anisotropy ($D$) in the case of
axial transition, the temperature variation of dynamic order parameter component $Q_z$ was studied for different
values of $D$. Figure \ref{hdt12}  shows the temperature variation of $Q_z$ for different values of $D$. Here, like the
case of off-axial transition \cite{ijmpc} the transition temperature increases as the strength of anisotropy increases.  
It is observed that the axial transition is discontinuous for lower
values of anisotropy strength $D$ (i.e., 0.5, 2.5 etc.) and it becomes continuous for higher values of $D$ (i.e., 5.0, 15.0 etc.). 
In the Ising limit ($D \rightarrow \infty$)
the axial transition is also shown in the same figure for $D = 400$). This choice of the value of $D (=400)$ is not arbitrary. In the case
of equilibrium transition it was checked by MC simulation that the value 
of the magnetisation at any temperature (in the ferromagnetic region)
becomes very close
to that obtained in the Ising model, at that temperature, if the strength of anisotropy is chosen above 300.

The temperature variations of dynamic order parameter component $Q_z$ in the
case of off-axial transition was also studied and shown in figure \ref{hdt13} for different values of $D$.
This shows that the transition temperature increases as $D$ increases. Here, the
transition is continuous for all values of strength of anisotropy $D$. 
The transition for $D$ = 400 ($D \rightarrow \infty$ limit)
was compared with that in the case of Ising model. This shows that both are continuous and occur at the same point
($T \approx 4.5$) which is very close to the Monte Carlo results of equilibrium ferro-para transition 
temperature ($T_c \simeq 4.511$) \cite{book} in 3-dimensional Ising model.

The nonequilibrium dynamical phase transition in the uniaxially anisotropic Heisenberg model, 
in presence of a magnetic field, which  oscillates sinusoidally in time,
is studied by Monte Carlo simulation.
Two cases were studied in this paper. (i) magnetic field, oscillates sinusoidally in time is applied
only along the direction of anisotropy, (ii) magnetic field applied only along the direction perpendicular
to the direction of anisotropy. The transition observed in the first case is named axial and that
corresponding to the second case is called off-axial. A comparative study between axial and off-axial
transition is reported in this paper. Three important aspects are considered here. (a) symmetry
breaking, (b) the order of the transition and (c) the transition in the infinitely anisotropic limit. 

A dynamic symmetry breaking is observed with this dynamic transition. In the case of axial transition
the dynamic transition occured as the temperature decreases from a symmetric to an asymmetric phase,
whereas, in off-axial case this symmetry breaking takes place from a 'marginally symmetric' to an asymmetric
phase. The reason behind it is that as follows: in the case of axial transition by the application of axial
field (oscillates sinusoidally in time) there is a chance that the spin component along the z-direction
may be reversed in opposite direction which would lead to sufficiently wide and symmetric $m_z-h_z$ loop.
But in the case of off-axial transition it is not possible to reverse 
the $z$-component of spin by applying 
a field (oscillates sinusoidally in time) perpendicular to the direction of uniaxial anisotropy. In
this case the value of the $z$-component of magnetisation $m_z$ is almost zero. As a result the 
$m_z-h_x$ loop lies on $h_x = 0$ axis and hence the loop is marginally symmetric.   

In both cases (axial and off-axial) the transition temperature increases as the strength of anisotropy
increases provided the amplitude of the applied field remains same. The strength of anisotropy tries to align 
the spin vector along the direction of anisotropy. So,
as the strength increases it becomes harder to break the symmetry and consequently more thermal fluctuation
is required to break the symmetry. As a result, the transition temperature increases as the strength of anisotropy
increases. But the difference is the nature of transition. In the case of axial field
the transition is discontinuous for lower values of anisotropy and it becomes continuous for higher values
of anisotropy. The reason behind it is, that the axial transition occurs in presence of axial field which reverses
the $z$-component of magnetisation. So, in lower values of anisotropy the spin vector becomes comparatively more flexible
and the transition occurs mechanically in presence of axial field at lower temperature and it is discontinuous. 
As the anisotropy increases the effect of axial field (of same value) becomes weak and the transition is driven by thermal 
fluctuations and the transition is continuous. In the case of off-axial transition, the off-axial field cannot 
reverse the $z$-component of magnetisation. But as the value of off-axial field increases, the value of $x$-component 
of magnetisation increases at the cost of $z$-component of magnetisation. The transition is driven by thermal 
fluctuations and continuous.  

What will be the situation in the limit of infinite strength of anisotropy ? In the case of axial transition it was observed that
the transition temperature for infinitely anisotropic Heisenberg model differs from that obtained in an Ising model.
Although the equilibrium transitions in the infinitely anisotropic Heisenberg model and in the Ising model gives the same
transition temperature, the nonequilibrium transition temperatures in those two cases are not same. 
Since the magnetic field is applied in $z$-direction, oscillating sinusoidally in time, keeps the system always away from
the equilibrium, the system does not become an Ising system even in the infinitely anisotropy limit. As a result, 
the dynamic transition temperature in the infinitely
anisotropic Heisenberg model cannot be same for that obtained in the Ising model.
But in the case of
off-axial transition, the transition temperatures in the infinitely anisotropic Heisenberg model and in the Ising model becomes   
exactly equal. The reason behind it is that as follows: in the case of off-axial transition the field is applied perpendicular to the
direction of anisotropy. The effect of axial field, oscillating sinusoidally in time, has no effect in the infinitely anisotropic
limit. Though the magnetic field is applied in the $x$-direction, oscillates sinusoidally in time, the infinite anisotropic
Heisenberg model becomes an Ising model in statistical and thermal equilibrium.  
Hence, the infinitely anisotropic Heisenberg model in presence of off-axial field maps into the Ising model in zero field.
That is why the nonequilibrium transition in the infinitely anisotropic Heisenberg model in presence of off-axial field
and the Ising model (in zero external field) give the same result.

One important point may be noted here regarding the iclassical dynamics chosen in this simulation. Since, the spin component
does not commute with the Heisenberg Hamiltonian the spin component has an intrinsic quantum dynamics. Considering this
intrinsic quantum dynamics there was a study \cite{land} about structure factor and transport properties in XY- model.
However, in this paper, the motivation is to study the nonequilibrium phase transition driven by thermal fluctuations. 
To study this, one should choose the dynamics which arises due to the interaction with thermal bath.
Since the objective is different, in this paper, the classical
dynamics chosen here (which arises solely due to the interaction with thermal bath), is Metropolis dynamics. The 
quantum 
effect of intrinsic
spin dynamics is not taken into account.

\vskip 0.5 cm

\noindent {\it (v) Multiple  dynamic transition}

In this case the MC study was done \cite{ma7} for a polarised magnetic field having the form: ${\vec h} = ih_x+jh_y+kh_z=
ih_{0x}{\rm cos}(\omega t)+kh_{0z}{\rm sin}(\omega t)$. One can readily check that $h_x = h_{0x}{\rm cos}(\omega t)$ and
$h_z = h_{0z}{\rm sin}(\omega t)$ yield, after the elimination of time 

\begin{equation}
{{h_x^2} \over {h_{0x}^2}} +{{h_z^2} \over {h_{0z}^2}} = 1
\label{pol}
\end{equation}

The simulational study is done for a simple cubic lattice of linear size $L = 20$. 
The total (vector) dynamic order parameter can be expressed as
$\vec Q = iQ_x + jQ_y + kQ_z$. The instantaneous energy 
$e(t) = -J \sum_{<ij>}\vec S_i \cdot \vec S_j - D \sum_i (S_i^z)^2 - \vec h \cdot \sum_i \vec S_i$ is also calculated.
The time-averaged instantaneous energy is $E =  \oint e(t) dt/\tau$. The rate of change of $E$ with respect to
the temperature $T$ is defined as dynamic specific heat $C ( = dE/dT)$ \cite{ma1}. The dynamic specific heat $C$
is calculated from energy $E$, just by calculating the derivative using the three-point central difference 
formula, given below.

\begin{equation}
C = {{dE} \over {dT}} = {{E(T+{\delta T})-E(T-{\delta T})} \over {2\delta T}}
\end{equation}
For the elliptically polarised (equation \ref{pol}) magnetic field, where the resultant field lies in $x-z$ plane, the amplitudes
of fields are taken as $h_{0x} = 0.3$ and $h_{0z} = 1.0$ and the frequency $f$ = 0.02. 
The strength of uniaxial anisotropy is taken $D = 0.2$.
This value of $D$ is obtained by rigorously searching to have these interesting results and kept constant throughout
the study. However, there must be variations in transition points depending on the values of $D$. 
It is observed that for higher values
of $D$ the multiple transition phenomenon disappears. The values of field amplitudes and frequency are also obtained
by searching.

The temperature variations of the dynamic order parameter components ($Q_x, Q_y, Q_z$) are studied and the results
are depicted in Fig.\ref{hdt14}(a). As the system is cooled down, from a high temperature disordered ($\vec Q = 0$) phase, 
it was observed that, first the system undergoes a transition
from dynamically disordered ($\vec Q$ = 0) to a dynamically {\it Y-ordered} (only $Q_y \ne 0$) phase. 
This may be called as the first phase ($P_1$) and the transition temperature is $T_{c1}$.
This phase can be characterised as $P_1$: ($Q_x = 0$, $Q_y \ne 0$, $Q_z = 0$).
Here, the resultant vector of elliptically polarised magnetic field
lies in $x-z$ plane and the dynamic ordering occurs along y-direction. So, this is clearly an off-axial transition \cite{ijmpc}.
In the case of this type of off-axial transition the dynamical symmetry (in any direction; y-direction here) 
is broken by the application of
the magnetic field in the perpendicular direction (lies in the $x-z$ plane here).
As the system cools down, it retains this particular dynamically ordered phase ($P_1$)
 over a considerable range of temperatures. As the temperature
decreases further, a second transion was observed. Here, the system becomes dynamically ordered both in X- and Z-directions at the
cost of Y-ordering. In this new dynamic phase,$P_2$: ($Q_x \ne 0$, $Q_y = 0$, $Q_z \ne 0$). 
In this phase the dynamical ordering is planar (lies on $x-z$ plane). 
The ordering occurs in the same plane on which the field vector lies.
This transition is axial \cite{ijmpc}. This phase may be called the second
phase ($P_2$) and the transition (from first phase to the second phase) temperature is $T_{c2}$. 
As the temperature decreases further, the $x$- and $z$-ordering increases.
At some lower temperature, a third transition was observed, 
from where the $x$-ordering starts to decrease and only $z$-ordering starts to increase quite rapidly.
This third phase can be designated as $P_3$: ($Q_x \ne  0$, $Q_y = 0$, $Q_z \ne 0$).
Although the characterisation of $P_2$ and $P_3$, in terms of the values of dynamic order parameter components, looks similar 
there exists an important difference between these two phases. In the phase $P_2$, both $Q_x$ and $Q_z$ increases as the tempereture
decreases but in the phase $P_3$, $Q_x$ decreases as the temperature decreases (see Fig.\ref{hdt14}(a)). So these two phases 
$P_2$ and $P_3$ distinctly
differ from each other.
In this phase the dynamical ordering is also axial (along $z$-axis or anisotropy axis).
The system continues to increase the dynamical $z$-ordering as the temperature decreases further. 
The low temperature phase is only dynamically $z$-ordered. That means the systems orders dynamically (only $Q_z \ne 0$) along the
$z$-direction (direction of anisotropy) only at very low temperatures. Zero temperature dynamic phase (for such polarised field)
can be characterised as $Q_x = 0$, $Q_y = 0$ and $Q_z = 1.0$.  

To detect the dynamic transitions and to find the transition temperatures the temperature variation of the energy
$E$ is plotted in Fig.\ref{hdt14}(b). From this figure it is clear that there are three dynamic transitions occur in this case.
The transition points are the inflexion points in $E-T$ curve. The temperature derivative of the energy $E$ is the
dynamic specific heat $C$. The temperature variation of $C$ is shown in Fig.\ref{hdt14}(c). The three dynamic transitions
are very clearly shown by three peaks of the specific heat plotted against the temperature $T$. From this figure
the transition temperatures are calculated (from the peak positions of $C-T$ curve). First transition (right peak)
occurs around $T_{c1}=1.22$, the second transition (middle peak) occurs at $T_{c2}=0.94$ and the third (left peak) transition
occurs around $T_{c3}=0.86$.

This study was further extended for other values of $h_{0x}$ keeping other parameters fixed. It was found that this
three transitions senario disappears for higher values of $h_{0x}$. For example, for $h_{0x} = 0.9$, the second phase
$P_2$ disappears. In this case, the $C-T$ curve shows two peaks. It was also observed that for $h_{0x} = 0.2$, 
$h_{0z} = 0.2$ (keeping all other parameter fixed) the system shows single transition and only dynamically orders along
$z$-direction.

In the present study, the external time-dependent magnetic field was taken elliptically polarised where the resultant
field vector rotates on $x-z$ plane. For the lower values of anisotropy and a specific range of the values of field
amplitudes the system undergoes multiple dynamic phase transitions. Here, three distinct phases are identified. In this paper,
this observation is just briefly reported. 
This multiple dynamic phase transition in anisotropic Heisenberg ferromagnet in the presence of elliptically polarised field, is
observed here by Monte Carlo simulation. An alternative method, to check this phenomenon, may be to use Landau-Lifshitz-Gilbert
equation of motion \cite{uli} with Langevin dynamics. Another important thing should be mentioned here regarding the possible explanation
of multiple dynamic phase transitions (axial and off-axial transitions) observed in the anisotropic Heisenberg model. One
possible reason may be the coherent rotation of spins. In contrast,  the dynamic phase transition
in the Ising model can be explained simply by spin reversal and nucleation \cite{stauffer}.
But to know the responsible mechanism behind the multiple dynamic phase transitions, observed in anisotropic
Heisenberg ferromagnets in the presence of polarised magnetic field, details investigations are required.

The variation of the dynamic phase boundaries with frequency and the
strength of anisotropy is quite interesting to be studied. This study also indicates that the system will show a very rich
phase diagram with multicritical behaviour. The finite size analysis is also necessary in order to distinguish the crossover
effects from the true phase transitions. This requires a huge computational task which will take much time. This work is in progress 
\cite{ma8} and the details will be reported later.

In the context of multiple dyanmic transition in anisotropic Heisenberg ferromagnet it should be mentioned here that a recent
study \cite{jang2} of anisotropic Heisenberg thin ferromagnetic film shows a double dynamic phase transition for surface and
bulk order parameter. The Hamiltonian for the classical Heisenberg ferromagnet with a bilinear exchange anisotropy $\lambda$,
in presence of competing surface fields as well as pulsed oscillatory fields, was taken \cite{jang2} as
\begin{equation}
H = -J \Sigma_{<ij>}[(1-\lambda)(S_i^xS_j^x + S_i^yS_j^y) + S_i^zS_j^z)]
-\Sigma_{i \epsilon {\rm surface1}} {\vec H_1} \cdot {\vec S_i} 
-\Sigma_{i \epsilon {\rm surfaceD}} {\vec H_D} \cdot {\vec S_i} - H(T)\Sigma_i S_i^z 
\end{equation} 
where ${\vec H_1}$ and ${\vec H_D}$ are the static applied surface fields and the time-dependent field $H(t)$ was taken to
have a pulsed form with
\begin{equation}
H(t)=
\left\{ 
\begin{array}{rcl}
-H_0, {{2(k-1)\pi} \over {\omega}} < t \leq  {{(2k-1)\pi} \over {\omega}}\\
H_0, {{2(k-1)\pi} \over {\omega}} < t \leq  {{2k\pi} \over {\omega}}

\end{array}\right.
\label{pof}
\end{equation}
Where $h_0$ is the amplitude and $\omega$ is the angular frequency of the oscillatory external field and
$k$ is an integer ($k = 1,2,3....$) representing the number of periods of the pulsed oscillatory external
field. The model film was taken \cite{jang2} a simple lattice of size $L\times L\times D$. 
The system is subject to competing applied surface fields in layers
$n = 1$ and $n = D$ of the film with ${\vec H_1} = h {\hat z} \delta_{i1}$ and ${\vec H_D} = -h {\hat z} \delta{iD}$.
The Monte Carlo study was done \cite {jang2} using Metropolis algorithm for $D=12$ and $L=32$. They calculated
the surface and bulk order parameter $<Q^{surface}>$ and $<Q^{bulk}>$ respectively and studied as a function of temperature.
From Fig. \ref{hdt15} it is clear that the critical temperature for the dynamic phase transition (DPT) in the surface layers is not the same
as that for bulk of the film. A double DPT was observed \cite{jang2} for a anisotropic Heisenberg film for
competing surface fields and pulsed oscillatory fields.

The off-axial\cite{ijmpc} and multiple DPT \cite{ma7,jang2} in anisotropic 
Heisenberg ferromagnet shows quite interesting
DPT which was not observed in the Ising model \cite{rev} and a rich phase diagram is expected here \cite{ma8}.   

\vskip 0.5 cm

\noindent {\large {\bf 5. Experimental evidences of dynamic phase transitions:}}

Several experimental works \cite{exp0,exp1,exp2} were performed to investigate the hysteretic responses as well
as the DPT in ferromagnetic samples.
In a recent experiment, Jiang et al \cite{exp2} studied the
frequency-dependent hysteresis of epitaxially 
grown ultrathin (2 to 6 monolayer
thick) Co films on a Cu(001) surface at room temperature. 
The films have
strong uniaxial magnetisation with two ferromagnetic phases of
opposite spin orientations. This  magnetic 
anisotropy  makes it appropriate to represent the system
by an Ising-like model.
The external
magnetic field $h(t)$ on the system was driven sinusoidally in the 
frequency ($f = \omega /2\pi$) range 0.1 to 500 Hz and in the amplitude
($h_0$) range 1 to 180 Oe. Here of course the time-varying current or
the magnetic field induces an eddy current in the core, 
which results in a 
counter-field reducing the effective magnitude of the applied field. 
The surface magneto-optical Kerr effect technique 
was used to measure the
response magnetisation $m(t)$. A typical variation of the 
loop area $A$ with the driving frequency $f$, at  room temperature
and at fixed external field amplitude $h_0$, is shown in Fig. \ref{expt}(a). 
Fig. \ref{expt}(b) shows clearly a signature of DPT and
corresponding symmetry breaking. Here, the dynamic order parameter $Q$ is
plotted against the field amplitude $h_0$ at a fixed frequency and emperature.
The inset of Fig. \ref{expt}(b) shows that for lower values of field amplitude
i.e., 12.0 Oe (left inset) the phase is dynamically ordered ($Q \neq 0$) and
asymmetric, and for higher values of field amplitude i.e., 48.1 Oe, the $m-h$
loop is symmetric and the phase is dynamically disordered ($Q = 0$).
This experimental observation supports the theoretical results of DPT
and dynamical symmetry breaking. However, an experimental study of
the entire phase boundary is yet to be done. 

\vskip 1cm

\noindent {\large {\bf 6. Summary:}}

     The DPT in model ferromagnetic systems (Ising and Heisenberg) in
the presence of sinusoidally oscillating magnetic field is reviewed here. This nonequilibrium DPT
is a prototype of nonequilibrium phase transition. The kind of nonequilibrium phase transition
discussed above is observed very recently. 
Here, only the observations are reviewed. The detailed mechanism behind
this type of nonequilibrium transition is not yet known clearly. In Ising models, the mechanism 
was tried to be understood in view of nucleation \cite{stauffer}. However, in the Heisenberg model the axial, off-axial \cite{ijmpc}
and very recently observed multiple DPT \cite{ma7} are just observed and the mechanisms responsible
for those transition are not yet known. The coherent spin rotation \cite{uli} may be a possible reason for this.
The experimental observations of the DPT \cite{exp2} are only made for Ising like (highly anisotropic) systems.  
Experimental studies are required to observe the special DPT in the Heisenberg
model. 

\vskip 1 cm

\noindent{\bf {Acknowledgments:}} The Library facility provided by Saha Institute of Nuclear Physics, Calcutta
is gratefully acknowledged. The author thankfully acknowledges collaborations with B. K.  Chakrabarti and D. Stauffer.
He would also like to thank Pratip Bhattacharyya for helping him to collect several important references, Sanjit
K. Das for preparing the electronic files of few figures by using scanner and A. Chatterjee for bringing Ref.
\cite{epl} into his notice.

\newpage

\begin{figure}
\begin{center}{\bf Figure Captions}\end{center}

\caption
{Schematic time variation
of the response magnetisation $m(t)$ compared to that of 
the oscillating field $h(t)$ for different values of frequency
$\omega$ and amplitude $h_0$
of the oscillating field and temperature $T$ of the system. 
The results are in fact actual Monte Carlo simulation
results for an Ising model on a square lattice 
 with the values for $h_0$ and $T$ as indicated in
the Figures. The Figures on the right hand side  show the 
corresponding $m-h$ loops. The values for loop area $A = \oint m dh$ and the dynamic
order parameter $Q$ are also indicated in these figures. As one
can see, the first two cases correspond to $Q =$ 0, while the
other two correspond to dynamically broken symmetric phase (with
$Q \ne$ 0). The first figure and the last correspond to $A \simeq$
0, while the middle two correspond to nonvanishing $A$.}
\label{dpt}
\end{figure}

\begin{figure}
\caption  {Phase diagrams in the $h_0$-$T$ plane for various
values of $\omega$ gives the functional form of the transition 
temperature  $T_d(h_0,\omega)$  for  the  dynamic  
phase
transition: Monte Carlo results {\bf (a)} for system sizes $L = 100$ in $d = 2$, and 
{\bf (b)} for $L$ = 20 in $d$ = 3. 
Below $T_d(h_0,\omega)$, $Q$ acquires a nonzero value in
F phase  and $Q$ = 0 in P  phase.  Different  symbols  denote  different
phase  boundary  lines  corresponding  to  
different  frequencies ($\omega$):
($\Box$) $\omega$ =
0.418, ($\triangle$) $\omega$ = 0.208,
($\diamond$) $\omega$ = 0.104 in {\bf (a)}; and ($\diamond$) $\omega$
= 0.418, ($\Box$) $\omega $ = 0.202,
($\circ$) $\omega$ = 0.104 in {\bf (b)}. The locations of
the tricritical points (TCP) are indicated by the circle. 
The insets show the
nature of the  transition  just  above (I: $h_0$ = 2.2 and 4.4 in
{\bf (a)}  and {\bf (b)} respectively) and below (II: 
$h_0$ = 1.8 and 3.6 in {\bf (a)} and {\bf (b)} respectively)
the  tricritical  points
along the phase boundaries.}
\label{ht}
\end{figure}

\begin{figure}
\caption {Monte Carlo results of the temperature ($T$) variation of
'relaxation' time ($\Gamma$) for two different values of field amplitudes
($h_0$):the bullet represents $h_0$ = 1.0 and the diamond represents 
$h_0$ = 0.5. Solid lines show the temperature ($T$) variations of dynamic
order parameter $Q$. Inset shows a typical 'relaxation' of $Q$ plotted 
against the number of cycles ($n$). The solid line is the best fit
exponential form of the data obtained from MC simulation. Here, $L$ = 100,
$\omega = 2 \pi \times 0.04$.}
\label{relax}
\end{figure}

\begin{figure}
\caption {Monte Carlo results of the temperature variations of
$C_{tot}$ for two different values of $h_0$: the filled square represents
$h_0$ = 0.8 and the filled triangle represents $h_0$ = 0.4. Solid lines
represent the temperature variations of $Q$. Inset shows the temperature
variations of $E_{tot}$ for two different values of $h_0$:(I) $h_0$ = 0.8
and (II) $h_0$ = 0.4. Here, $L$ = 100, $\omega = 2 \pi \times 0.01$.}
\label{spht}
\end{figure}

\begin{figure}
\caption {(a) The histograms of the normalized distributions of the dynamic order parameter $Q$
for different temperatures ($T = 0.20J/K_B$, $0.28J/K_B$, $0.30J/K_B$ and $0.40J/K_B$) 
and for the fixed value of field amplitude $h_0$.
All the figures are plotted in the same scales.
(b) The normalized distributions of the dynamic order parameter $Q$ ( in the 2nd order
and close to the transition region ) for three different temperatures ($T = 1.48J/K_B$, 
$1.50J/K_B$, $1.55J/K_B$) and fixed field amplitude
$h_0 = 0.3J$.} 
\label{qdst}
\end{figure}

\begin{figure}
\caption {The histogram of normalized ($\int P_r(\tau_r) d\tau_r = 1$)
distribution ($P_r(\tau_r)$) of the residence
time ($\tau_r$). 
The dotted line is the exponential best fit of the envelope
of the distribution.}
\label{sr}
\end{figure}

\begin{figure}
\caption {Dynamic transition due to randomly
varing fields in time.
{\bf (a, b)} Typical time variation of magnetisation $m(t)$ 
compared to that of the stochastically varying field $h(t)$
in a Monte Carlo study in $d = 2$, 
with
$ L = 100, T = 1.7$:  $h_0 = $ 1.0 for 
{\bf (a)}  and $h_0 =$ 3.0 for {\bf (b)}.
{\bf (c)} The  corresponding dynamic transition phase boundary
(separating the regions with average magnetisations $Q$  zero from
nonzero) 
in the field width ($h_0$) - temperature ($T$) plane. The 
data points are obtained using both sequential updating ($\diamond$)
and random updating ($\bullet$) in the Monte Carlo simulation.}
\label{rnd}
\end{figure}

\begin{figure}
\caption {(a) The phase boundary (of the dynamic transition) 
in $w-h_0$ plane. The tricritical point (TCP) lies within the encircled
region. The boundary of the circle is the uncertainty associated in locating
the TCP, (b) two typical transitions just below and above the tricritical
point which show the different natures (discontinuous/continuous)
 of the transitions.}
\label{dpt0}
\end{figure}

\begin{figure}
\caption {Symmetry breaking in axial and off-axial transitions.
The plot of instantaneous magnetization components against the instantaneous
field components. 
(a) $m_x(t)-h_z(t)$ and $m_z(t)-h_z(t)$ loops for $D = 2.5$, $h_z^0 = 0.5$ and $T = 2.2$,
(b) $m_x(t)-h_z(t)$ and $m_z(t)-h_z(t)$ loops for $D = 2.5$, $h_z^0 = 0.5$ and $T = 1.0$,
(c) $m_x(t)-h_x(t)$ and $m_z(t)-h_x(t)$ loops for $D = 0.5$, $h_x^0 = 0.5$ and $T = 1.8$ and
(d) $m_x(t)-h_x(t)$ and $m_z(t)-h_x(t)$ loops for $D = 0.5$, $h_x^0 = 0.5$ and $T = 0.6$.}
\label{hdt11}
\end{figure}

\begin{figure}
\caption {The axial dynamic transitions. 
Temperature ($T$) variations of dynamic order parameter components $Q_z$
for different values of anisotropy strength ($D$) represented by different symbols. $D = 0.5 (\Diamond)$,
$D = 2.5 (+)$, $D = 5.0 (\Box)$, $D = 15.0 (\times)$ and $D = 400.0 (\triangle)$. In all these cases
for the ${\it axial}$ transitions $h_z^0 = 0.5$. 
The data for the temperature variation of dynamic order parameter in 
the Ising model (for $h_z^0 = 0.5$ and $f = 0.001$) are represented by $\star$. Continuous lines in all 
cases are just connecting the 
data points.}
\label{hdt12}
\end{figure}

\begin{figure}
\caption {The off-axial dynamic transitions. 
Temperature ($T$) variations of dynamic order parameter components $Q_z$
for different values of anisotropy strength ($D$) represented by different symbols. $D = 0.5 (\Diamond)$,
$D = 2.5 (+)$, $D = 5.0 (\Box)$, $D = 15.0 (\times)$ and $D = 400.0 (\triangle)$. In all these cases
for ${\it off-axial}$ transitions $h_x^0 = 0.5$. 
The data for the zero-field ferro-para
equilibrium Ising transition are represented by $\star$. Continuous lines in all 
cases are just connecting the 
data points.}
\label{hdt13}
\end{figure}

\begin{figure}

\caption {(a) The temperature variations of the components of dynamic order parameters. Different
symbols represent different components. $Q_x$ (diamond), $Q_y$ (circle) and $Q_z$ (bullet). This
diagram is for $D = 0.2$ and for elliptically polarised field where $h_{0x}=0.3$ and $h_{0z}=1.0$. 
The size of the errorbars of $Q_x$, $Q_y$ and $Q_z$ close to the transition points is of the order of 0.02 and 
that at low temperature (e.g., below $T = 0.5$) is around 0.003. (b) The temperature variation of the dynamic 
energy ($E$) for $D = 0.2$, $h_{0x} = 0.3$ and $h_{0z} = 1.0$. The vertical arrows represent the transition points.
(c) The temperature variation of dynamic specific heat ($C = {dE \over dT}$) for $D = 0.2$, $h_{0x} = 0.3$
and $h_{0z} = 1.0$. Vertical arrows show the peaks and the transition points.}

\label{hdt14}
\end{figure}

\begin{figure}
\caption {Surface order parameter $<Q^{surface}>$ and bulk order parameter $<Q^{bulk}>$ 
for the film, plotted as a function
of temperature ($T^*$) for the value of pulsed oscillatory field $H_0 = 0.3$.}
[After Jang et al, Phys. Rev. B 67 (2003) 094411]
\label{hdt15}
\end{figure}

\begin{figure}
\caption {Experimental results for the dynamic hysteresis
loop area $A$ and the dynamic order parameter $Q$ \cite{exp2}.
 {\bf (a)} The  results for the
loop area $A$ as a function of frequency $f$
is plotted at a fixed ac current of 0.4 
Amp. The direction of the magnetic field
is  parallel to the film plane. 
The insets show plots of $m-h$ loops for the following particular values of
the field amplitudes $h_0$: (i) $h_0 = 48.0$ Oe (top inset) and (ii) $h_0 = 63.0$ Oe (bottom inset).
{\bf (b)} The dynamic order parameter $Q$, i.e,
the average magnetisation over a cycle, is plotted 
against the field amplitude at a fixed frequency $f$ = 4 Hz.
The insets show plots of $m-h$ loops for the following particular values of
the field amplitudes $h_0$: (i) $h_0 = 48.1$ Oe (right inset) and (ii) $h_0 = 12.0$ Oe (left inset).
[After Q. Jiang et al. Phys. Rev. B. 52 (1995) 14911] }
\label{expt}
\end{figure}

\vskip 1 cm

\begin{center} {\large {\bf References}}\end{center}

\begin{enumerate}

\bibitem{rev} B. K. Chakrabarti and M. Acharyya, {\it Rev. Mod. Phys.}, {\bf 71}, (1999), 847

\bibitem{tom} T. Tome and M. J. de Oliveira, {\it Phys. Rev.} A {\bf 41}, (1990) 4251

\bibitem{rao} M. Rao, H. R. Krishnamurthy and R. Pandit, {\it Phys. Rev. B}, {\bf 42} (1990) 856

\bibitem{dd} D. Dhar and P. B. Thomas, {\it J. Phys. A} {\bf 25} (1992) 4967

\bibitem{lo} W. S. Lo and Pelcovits, {\it Phys. Rev. A} {\bf 42} (1990) 7471

\bibitem{ac} M. Acharyya and B. K. Chakrabarti, {\it Phys. Rev.} B {\bf 52} (1995) 6550; see also M. Acharyya
and B. K. Chakrabarti in {\it Annual Reviews of Computational Physics}, Vol-I, Edited by D. Stauffer,
(World Scientific, Singapore) (1994), p 107.

\bibitem{ma1} M. Acharyya, {\it Phys. Rev.} E, {\bf 56} (1997) 2407 

\bibitem{ma2} M. Acharyya, {\it Phys. Rev.} E, {\bf 56} (1997) 1234

\bibitem{ma3} M. Acharyya, {\it Phys. Rev.} E, {\bf 58} (1998) 179

\bibitem{ma4} M. Acharyya, {\it Phys. Rev.} E, {\bf 58} (1998) 174

\bibitem{ma5} M. Acharyya, {\it Phys. Rev.} E, {\bf 59} (1999) 218

\bibitem{ma6} M. Acharyya, {\it Physica A}, {\bf 253} (1998) 199

\bibitem{rik1} S. W. Sides, P. A. Rikvold and M. A. Novotny, {\it Phys. Rev. E} {\bf 57} (1998) 6512

\bibitem{rik2} S. W. Sides, P. A. Rikvold and M. A. Novotny, {\it Phys. Rev. Lett.}, {\bf 81} (1998) 834

\bibitem{rik3} G. Korniss, P. A. Rikvold and M. A. Novotny, {\it Phys. Rev. E}, {\bf 66} (2002) 056127

\bibitem{stauffer} M. Acharyya and D. Stauffer, {\it Eur. Phys. J. B}, {\bf 5}, (1998) 571


\bibitem{mc} D. Stauffer {\it et al.}, {\it Computer simulation and Computer Algebra}\\
 (Springer- Verlag, Heidelberg,
1989); K. Binder and D. W. Heermann,\\ {\it Monte Carlo simulation in Statistical Physics, Springer Series\\
 in Solid- State Sciences} (Springer, 1997) 

\bibitem{arnab} A. Chatterjee and B. K. Chakrabarti, 
{\it Phase Transitons}, {\bf 77} (2004) 581.  


\bibitem{sr} L. Gammaitoni, P. H\"anggi, P. Jung and F. Marchesoni, {\it Rev. Mod. Phys.}, {\bf 70}, 223 (1998).


\bibitem{binder} K. Binder, K. Vollmayr, H-P Deutsch, J. D. Reger and M. Scheucher,
 {\it Int. J. Mod. Phys.} C, {\bf 3}, 1025 (1992); 

\bibitem{haus} J. Hausman and P. Ruj$\acute{\bf a}$n, {\it Phys. Rev. Lett.} {\bf 79} (1997) 3339

\bibitem{shukla} D. Dhar, P. Shukla and J. P. Sethna, {\it J. Phys. A: Math \& Gen} {\bf 30} (1997) 5259

\bibitem{rfim} M. Acharyya, {\it Physica A} {\bf 252} (1998) 151




\bibitem{book} D. C. Mattis, {\it The theory of magnetism I: Statics and Dynamics, Springer Series in Solid- State 
Sciences}, Vol. {\bf 17} (Springer- Verlag, Berlin, 1988).

\bibitem{ijmpc} M. Acharyya, {\it Int. J. Mod. Phys.} C {\bf 14} (2003) 49; see also {\it Int. J. Mod. Phys.} C
{\bf 12} (2001) 709.

\bibitem{yasui} T. Yasui {\it et al.} {\it Phys. Rev.} E, {\bf 66} (2002) 036123; 
Errutum, {\it Phys. Rev. E}, {\bf 67} (2002) 019901(E)

\bibitem{epl} I. Junier and J. Kurchan, Europhys. Lett, {\bf 63} (2003) 674

\bibitem{ma7} M. Acharyya, {\it Phys. Rev.} E {\bf 69} (2004) 027105.

\bibitem{land} M. Krech and D. P. Landau, {\it Phys. Rev. B} {\bf 60} (1999) 3375

\bibitem{uli}  U. Nowak in {\it Annual Reviews of Computational Physics}, {\bf 9} Ed. D. Stauffer, 
World-Scientific, Singapore, (2001) p.105; D. Hinzke and U. Nowak, {\it Phys. Rev.} B {\bf 58} (1998) 265;

\bibitem{jang2} H. Jang, M. J. Grimson and C. K. Hall, {\it Phys. Rev.} B 67 (2003) 094411; See also
{\it Phys. Rev.} E 68 (2003) 046115

\bibitem{ma8} M. Acharyya, {\it in preparation}, (2005).

\bibitem{exp0} Y. L. He and G. -C. Wang, {\it Phys. Rev. Lett}, {\bf 70} (1993) 2336

\bibitem{exp1} J. S. Suen and J. L. Erskine, {\it Phys. Rev. Lett} {\bf 78} (1997) 3567; J. S. Suen,
M. H. Lee, G. Teeter and J. L. Erskine, {\it Phys. Rev. B}, {\bf 59} (1999) 4249

\bibitem{exp2} Q. Jiang, H. N. Yang and G. -C. Wang, {\it Phys. Rev. B} {\bf 52} (1995) 14911

\end{enumerate}
\end{document}